\documentclass[a4paper]{jpconf}
\usepackage{graphicx}
\usepackage{amsmath}
\usepackage{microtype}
\usepackage[utf8]{inputenc}
\begin{document}

\mbox{}\hfill LU TP 15-44\\
\mbox{}\hfill MCnet-15-30\\
\mbox{}\hfill October 2015\\
\vspace{-3\baselineskip}

\title{Monte Carlo event generation of photon-photon collisions at colliders}

\author{I Helenius}

\address{Department of Astronomy and Theoretical Physics, Lund University, S\"{o}lvegatan 14A,\\ SE-223 62 Lund, Sweden}

\ead{ilkka.helenius@thep.lu.se}

\begin{abstract}
In addition to being interesting in itself, the photon-photon interactions will be an inevitable background for the future electron-positron colliders. Thus to be able to quantify the potential of future electron-positron colliders it is important to have an accurate description of these collisions. Here we present our ongoing work to implement the photon-photon collisions in the \textsc{Pythia}~8 event generator. First we introduce photon PDFs in general and then discuss in more detail one particular set we have used in our studies. Then we will discuss how the parton-shower algorithm in \textsc{Pythia}~8 is modified in case of photon beams and how the beam remnants are constructed. Finally a brief outlook on future developments is given.
\end{abstract}

\section{Introduction}

The particle spectrum of Standard Model (SM) was completed by the discovery of Higgs boson at the first run of the LHC. To go beyond the SM physics there are currently no guiding principle that would tell where new physics should be found, or whether there will be any within the reach of feasible experiments. Thus it is unclear which kind of experiments would be optimal to deepen our understanding of nature. Currently the popular options are to build a linear $\mathrm{e^+}\mathrm{e^-}$ collider with $\mathcal{O}$(TeV) energy (ILC and CLIC) or to build a circular collider with electron/positron beams (FCC-ee at CERN and CepC at IHEP, China) as a first phase and later use the same tunnel for a hadronic collider (FCC-hh, SppC) as a second phase. In either case it seems that the next high-energy collider will be an $\mathrm{e^+}\mathrm{e^-}$ one.

The advantage of an $\mathrm{e^+}\mathrm{e^-}$ collider is that here the hard process takes the full energy. However, the electrons emit photons which can interact with other photons, creating a background for $\mathrm{e^+}\mathrm{e^-}$ interactions. To quantify the physics potential of the future collider experiments one should thus be able to simulate these additional $\gamma\gamma$ interactions with a good precision.

Here we present our current development efforts to model the $\gamma\gamma$ collisions with the \textsc{Pythia}~8 \cite{Sjostrand:2014zea} general purpose Monte Carlo event generator. Such collisions were implemented in the former \textsc{Pythia}~6 \cite{Sjostrand:2006za} generator, but in an increasingly convoluted form. Here our aim is for a clean start with a new framework that makes use of the recent developments in the event generation.

\section{Photon PDFs}

In leading order one can indentify two separate components for the interaction of a high-energy photon: it can interact as an unresolved particle but it can also fluctuate to a hadronic state with the same quantum numbers. The cross section in the former case can be computed from perturbative QCD (pQCD), but for the latter case we need parton distribution functions (PDFs) for photons to describe their partonic content. The PDFs are non-perturbative objects but they can be obtained through a global QCD analysis using DGLAP evolution equations.

\subsection{Evolution equations}

The scale evolution of photon PDFs is given by
\begin{equation}
\frac{\mathrm{\partial} f^{\gamma}_i(x,Q^2)}{\mathrm{\partial}\mathrm{log}(Q^2)} = \frac{\alpha_{\rm EM}}{2\pi}e_i^2 P_{i\gamma}(x) + \frac{\alpha_s(Q^2)}{2\pi} \sum_j \int_x^1\frac{\mathrm{d}z}{z}\, P_{ij}(z)\, f_j(x/z,Q^2),
\label{eq:gammaDGLAP}
\end{equation}
where $f^{\gamma}_i(x,Q^2)$ is the PDF for flavor $i$, $P_{ij}(z)$'s are the splitting functions for $j\rightarrow ik$ splittings and the sum runs over relevant parton flavors $j$. The difference to the ordinary DGLAP equation for hadrons is that now there exists an extra term $\alpha_{\rm EM}/(2\pi)\,e_i^2 P_{i\gamma}(x)$ which arises from $\gamma\rightarrow q\bar{q}$ splittings. In leading order (LO) $P_{i\gamma}(x)=3\,[x^2+(1-x)^2]$ for quarks and zero for gluons.

As there are more terms in the evolution equation, also the solution has richer structure than in case of hadrons. The solution of equation \ref{eq:gammaDGLAP} can be decomposed into two parts
\begin{equation}
f^{\gamma}_i(x,Q^2) = f^{\gamma, \mathrm{pl}}_i(x,Q^2) + f^{\gamma, \mathrm{had}}_i(x,Q^2),
\label{eq:hadpl}
\end{equation}
where the point-like part $f^{\gamma, \mathrm{pl}}_i(x,Q^2)$ is a solution of the full inhomogeneous differential equation arising from $\gamma\rightarrow q\bar{q}$ splittings, and the hadron-like part $f^{\gamma, \mathrm{had}}_i(x,Q^2)$ corresponds to a general solution of the homogeneous part of the equation. The former can be calculated using pQCD with appropriate boundary condition at the chosen initial scale $Q_0$, but for the latter part some non-perturbative input is required at $Q_0^2$, which needs to be fixed by data. Often one utilizes the vector meson dominance (VMD) model for the hadron-like part, so it can be written as a sum of light vector meson PDFs. In first analyses, e.g. in GRV \cite{Gluck:1991jc}, the shape of the hadron-like part was taken to be the same as for pions and only the normalization was fitted to data. In more recent analyses, e.g. in CJKL \cite{Cornet:2002iy}, also the shape have been fixed by data, typically using an ansatz of the form $f_i^{\gamma, \mathrm{had}}(x,Q_0^2) = N_i\,x^{a_i}\,(1-x)^{b_i}$.

\subsection{CJKL analysis}

There are several photon PDF analysis available nowadays that include all the available photon structure function data from LEP experiments, both at leading and next-to-leading order. Here we have chosen to use photon PDFs from CJKL analysis \cite{Cornet:2002iy}. One reason for this is that since \textsc{Pythia}~8 is a LO event generator one should use PDFs that are determined at the same order. Another useful feature in CJKL analysis is that it provides parametrizations for the hadron-like and point-like contributions separately, and the hadron-like part is further separated as valence and sea quark contributions. This information is useful for the beam-remnant handling, as will be discussed in section \ref{sec:beamRemnants}. An optimal analysis in this respect would separate also the partons created in $\gamma \rightarrow q\bar{q}$ splittings from the further QCD-evolution of this quark-antiquark pair to provide further information for the beam remnant construction. This was actually the case in SaSgam analysis \cite{Schuler:1995fk} but it precedes most of the LEP data, so a fit that includes also the most recent data was preferred.

\begin{figure}
\centering
\includegraphics[width=0.6\textwidth]{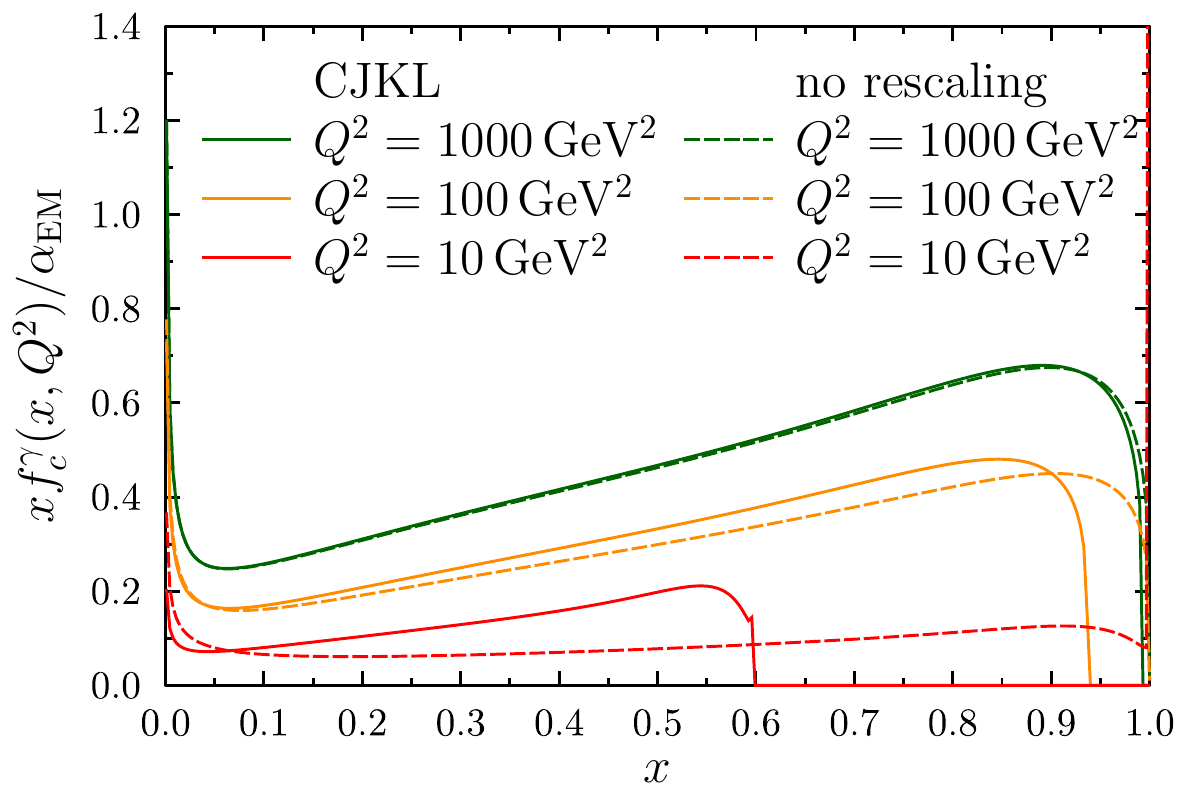}
\caption{Charm quark PDF from CJKL analysis with the rescaled $x$ (solid) and after the rescaling have been undone (dashed) for scales $Q^2=10\,\mathrm{GeV^2}$ (red), $Q^2=100\,\mathrm{GeV^2}$ (orange) and $Q^2=1000\,\mathrm{GeV^2}$ (green).}
\label{fig:UnReScaling}
\end{figure}

When performing a PDF analysis one needs to decide how to deal with heavy quark masses. The CJKL analysis has adopted so called ACOT($\chi$) scheme \cite{Tung:2001mv} in which the usual Bjorken-$x$ is replaced with a rescaling variable $\chi_h = x(1+4\,m^2_h/Q^2)$, where $m_h$ is the mass of heavy quark $h$, to obtain desired smooth vanishing of the heavy quark contribution near the mass threshold. The rescaling variable arises from the condition $W^2=Q^2(x^{-1}-1) > (2m_h)^2$, where $W$ is the invariant mass of the $\gamma\gamma$ pair. This condition, however, is specific for DIS kinematics and in $\gamma\gamma$ collision the limit for heavy quark production is not related to $Q^2$ but rather to $\sqrt{s}$, so we need to undo this rescaling. This is done by stretching the heavy quark PDFs to cover the whole $x$ region but keeping the integral over $x$ fixed. The procedure is illustrated in figure \ref{fig:UnReScaling}, which shows the charm quark PDFs with and without this rescaling for three different values of $Q^2$.

\section{Parton showers}

The partons taking part in the hard process of interest can emit additional partons before and after the hard interaction. The partons that are radiated before the interactions are typically referred as initial state radiation (ISR) and the partons radiated after the hard interactions as final state radiation (FSR). The parton shower generation is based on the same DGLAP equations as the PDFs. For FSR, where evolution of single particle is considered, the splitting probability at given scale $Q^2$ and $z$ is given simply by
\begin{equation}
\mathrm{d}\mathcal{P}_{a\rightarrow b c} = \frac{\mathrm{d}Q^2}{Q^2} \frac{\alpha_s}{2\pi} P_{a\rightarrow b c}(z) \, \mathrm{d}z.
\end{equation}
As FSR is deals only with individual partons there are no difference between photon and proton beams.

The splitting probability for the ISR can also be obtained from the DGLAP equation. However, as the shower here is reconstructed ``backwards in time'', starting at the hard process, one needs to consider conditional probability for a branching $a\rightarrow b c$ given a parton $b$ with known $x$ and $Q^2$. Thus the probability for one splitting becomes
\begin{equation}
\mathrm{d}\mathcal{P}_{a\rightarrow b c} = \frac{\mathrm{d}Q^2}{Q^2} \frac{x'f^{\gamma}_a(x',Q^2)}{xf^{\gamma}_b(x,Q^2)}\frac{\alpha_s}{2\pi}P_{a\rightarrow b c}(z)\, \mathrm{d}z + \frac{\mathrm{d}Q^2}{Q^2} \frac{\alpha_{\mathrm{EM}}}{2\pi}\frac{e_b^2\,P_{\gamma\rightarrow bc}(x)}{f^{\gamma}_b(x,Q^2)},
\end{equation}
where the latter term again corresponds to probability of finding the original beam photon during the evolution, and $x'=x/z$. 

The full shower can then be generated by evolving from a starting scale related to the hard process down to a minimum scale below which the hadronization occurs. During this $Q^2$-ordered evolution one needs to take into account the no-emission probabilities. This can be accomplished by multiplying the above splitting probabilities by Sudakov form factors. For further details of the \textsc{Pythia}~8 implementation in case of a hadron beam, see Ref.~\cite{Sjostrand:2004ef}. In case of a photon beam the $\gamma\rightarrow q\bar{q}$ splitting is added to the ISR algorithm. As the momentum of the beam photon is known, there is no need to sample the $z$, it is simply given by the $x$ of the daughter parton. To sample the $Q^2$, it is useful to notice that the scale evolution of the PDFs at given $x$ goes roughly as $\log{Q^2}$ and write
\begin{equation}
\frac{P_{\gamma\rightarrow bc}(x)}{f^{\gamma}_b(x,Q^2)} = \frac{c_b\log{(Q^2/Q_0^2)}}{f^{\gamma}_b(x,Q^2)}\frac{P_{\gamma\rightarrow bc}(x)}{c_b\log{(Q^2/Q_0^2)}},
\end{equation}
where $Q_0 = m_c$ for $\gamma \rightarrow c\bar{c}$ and $c_b$ is fixed so that $w=(c_b\log{(Q^2/Q_0^2)})/f^{\gamma}_b(x,Q^2) < 1$. Then one can use the part $P_{\gamma\rightarrow bc}(x)/(c_b\log{(Q^2/Q_0^2)})$ for $Q^2$ sampling, and correct the probability later with the weight $w$. At the moment $c_b$ is taken as a constant, but one could also introduce some $x$ dependence to increase the efficiency of the algorithm. A more detailed description of the ISR algorithm with photon beams will be given in a future publication \cite{HeleniusSjostrand}. 

\begin{figure}[htb]
\begin{minipage}[t]{0.485\linewidth}
\centering
\includegraphics[width=\textwidth]{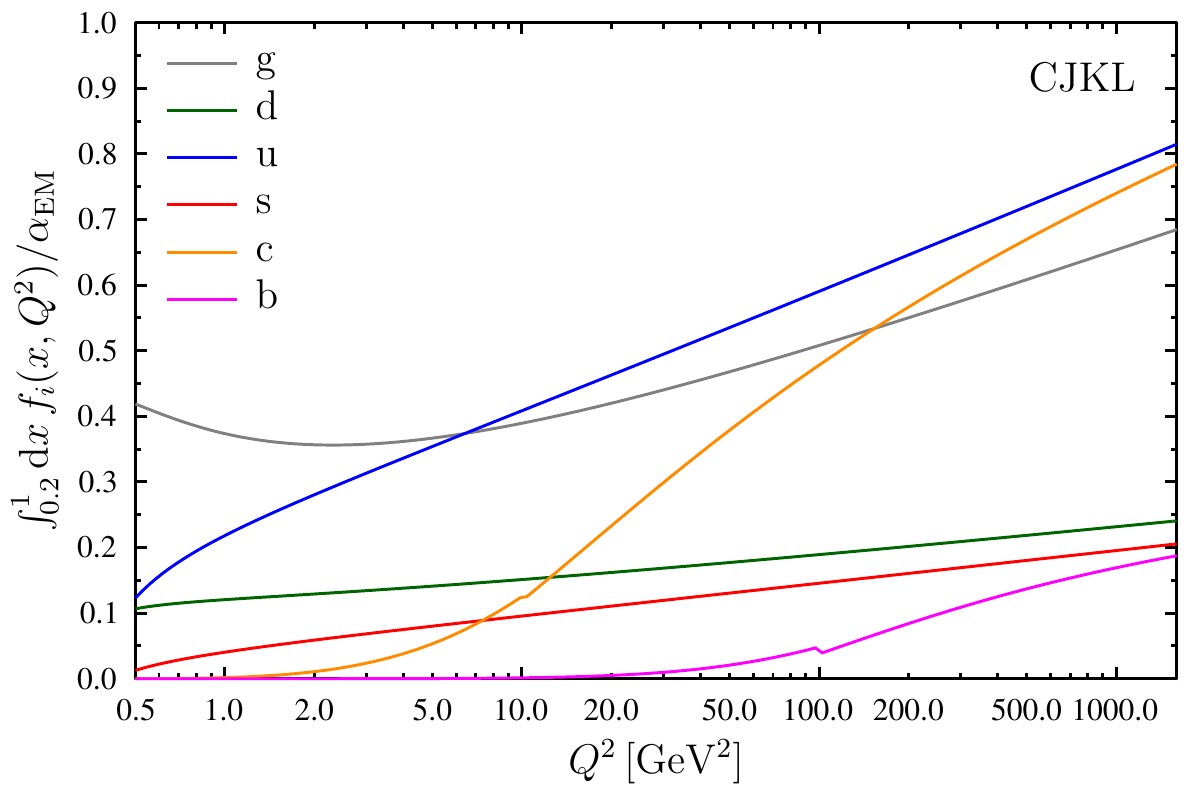}
\caption{The $x$-integrated PDFs from CJKL analysis for gluons (gray), d- (green), u- (blue), s- (red), c- (orange) and b- (magenta) quarks.}
\label{fig:fQ2int}
\end{minipage}
\hspace{0.01\linewidth}
\begin{minipage}[t]{0.485\linewidth}
\centering
\includegraphics[width=\textwidth]{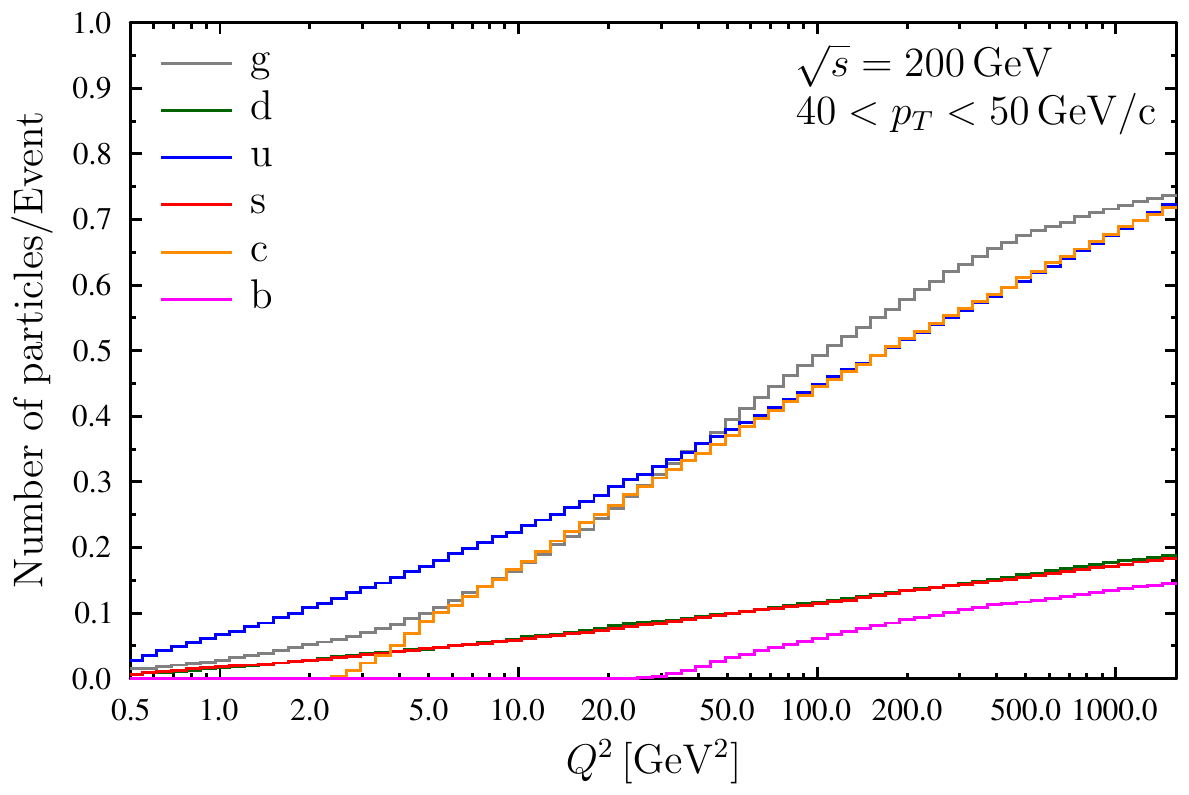}
\caption{Number of partons/event produced below scale $Q^2$ by the ISR algorithm for $\gamma\gamma$-collision at $\sqrt{s}=200\,\mathrm{GeV}$ in events with $40 < \hat{p}_T < 50\,\mathrm{GeV/c}$ for different parton flavors with the same color coding as in figure \ref{fig:fQ2int}.}
\label{fig:nPartQ2}
\end{minipage}
\end{figure}

As the backwards evolution (going from a larger scale towards a smaller one) of the ISR algorithm is based on the same evolution equations as the forward evolution (from smaller to larger scale) in PDFs these approaches can be compared to check whether the modified ISR algorithm works as expected. This comparison is shown in figures \ref{fig:fQ2int} and \ref{fig:nPartQ2} where the former shows the PDFs from CJKL fit for each parton flavor integrated over $0.2 < x < 1$ as a function of $Q^2$ and the latter shows the number of partons per event produced by the ISR algorithm below a certain scale for $\gamma\gamma$ collision at $\sqrt{s}=200\,\mathrm{GeV}$ for events with $40 < \hat{p}_T < 50\,\mathrm{GeV/c}$, where $\hat{p}_T$ is the transverse momentum of the hard partonic $2\rightarrow2$ process.
The main features seem to come out in a similar manner for the two approaches: heavy quarks are not produced below their mass threshold and in general parton decomposition look very similar. However, as not all the details are the same in these approaches, there are also some differences. First, the number of charm quarks from ISR behaves as $\sim\log Q^2$ but in the CJKL PDFs the number grows first very slowly, then even faster than for u-quarks, and finally levels to similar a $\log Q^2$-evolution as the u-quarks. These features arise due to the ACOT($\chi$) scheme used for heavy quark mass effects applied in the CJKL analysis. Second, the number of gluons at small values of $Q^2$ is clearly higher in the CJKL PDFs. The data used in the analyses does not provide much sensitivity for gluon PDFs, and the number of gluons at the initial scale is fixed by energy-momentum sum rule in the CJKL fit. This hints that the large number of gluons at small $Q^2$ is rather due to the choice of the free parameters in the analysis than being a feature in the data. The observed decrease of the number of gluons at $Q^2<2\,\mathrm{GeV^2}$ follows from the introduced $x>0.2$ cut as the DGLAP evolution shifts gluons to lower values of $x$.

\section{Beam Remnants}
\label{sec:beamRemnants}

In the case of proton beams the valence content is fixed and well defined. After the parton shower is generated the remnants can be constructed by first deciding whether the parton taken from the beam was a valence parton or not, using information in the PDFs, and then adding the correct number of parton such that flavor and total momentum in conserved. The procedure in case of \textsc{Pythia}~8 is described in detail in Ref.~\cite{Sjostrand:2004pf}.

For the photon beams there are some further complications. First of all, the valence content is not as well-defined. There are two components in the PDFs that can be interpreted as valence partons, however. The first is the valence contribution of the hadron-like PDF and the second is the part of the point-like PDF in which the partons originate from $\gamma \rightarrow q\bar{q}$ splittings. If all this information is available in the PDFs one can decide whether the parton taken from the beam was a valence parton or not, and construct the remnants accordingly. 

Partons in photon may carry a very large momentum fraction, so care needs to be taken in the hard-process and the parton-shower generation to ensure that there still is room left for the beam remnants. If ISR is not applied, the valence content need to be determined during the hard-process generation. Further, only processes where the remaining invariant mass is sufficient to accommodate the required quarks with their masses are allowed. The definitive constraint for the invariant mass can be obtained from the case where valence-type quarks from each beam interact with each other, and only one quark needs to be added to each beam remnant. If the potential primordial $p_T$ of the interacting quarks is neglected, the invariant mass left for the remnants is $W_{\rm rem} = \sqrt{s(1-x_1)(1-x_2)}$. The invariant mass for each remnant in this case is simply the mass of the remnant quark so the condition for kinematically possible hard process becomes
\begin{equation}
\sqrt{s(1-x_1)(1-x_2)} > m_j + m_k,
\label{eq:Wrem}
\end{equation}
where $m_j$ and $m_k$ are the masses of the remnant quarks in each beam. If the interacting partons are not valence-like, the constraint becomes somewhat more complicated.

When the ISR is applied, there is also a possibility that no beam remnants are needed, if the ISR ends up at the original beam photon for the given beam. Therefore three alternatives exist: remnants need to be constructed on both sides, on one side only, or at neither of the sides. The last case is the simplest as the parton-shower algorithm has already created all the required partons with correct momenta. If the beam photon is not found in either of the sides, the situation is very similar as with protons and the existing procedure can be used, where the sampled remnant momenta are balanced between the two sides to ensure the total momentum conservation. In the remaining case, where only one remnant needs to be constructed, the momentum cannot be balanced with the other side as the momentum of the beam photon is already fixed. In this case the momentum is balanced between the scattered partons and the remnants. One should also make sure that the parton-shower evolution does not end up in a situation where e.g. equation (\ref{eq:Wrem}) is violated, and the beam remnants cannot be constructed. The beam-remnant handling with the photon beams will be further discussed in the future article \cite{HeleniusSjostrand}.

\section{Summary}

We have been working on the implementation of $\gamma$$\gamma$ collisions in the \textsc{Pythia}~8 event generator. The current status is that we have included one set of photon PDFs which can be used to generate the hard process, and the ISR algorithm have been modified to include an additional splitting which corresponds to $\gamma \rightarrow q\bar{q}$ of the original beam photon. Also the beam-remnant handling is modified to work correctly with the photon beams with and without ISR. These developments will allow the generation of full $\gamma\gamma$ events including hadronization of the partons. The modifications will be added to a public version of \textsc{Pythia}~8 and a publication with a more detailed description of the new features is in preparation.

So far we have been considering only real photons, and only hard interactions, but in future we would like to take into account also the virtuality of the colliding photons and model the photon emissions from the colliding electrons. Another future development will be to include a possibility for soft processes and multiple partonic interactions in $\gamma$+$\gamma$ collisions.

\ack
Work have been supported by the MCnetITN FP7 Marie Curie Initial Training Network, contract PITN-GA-2012-315877.

\section*{References}

\bibliographystyle{iopart-num}
\bibliography{Photon2015Proceedings}

\end{document}